\def\br{{\bf r}}
\def\bk{{\bf k}}
\def\bE{{\bf E}}
\def\bR{{\bf R}}
\def\ekj{\hat{e}_{\bk j}}
\def\akj{a_{\bk j}}
\def\akjd{a_{\bk j}^\dagger}
\def\Fln{F_{\ell n}}
\def\Flm{F_{\ell m}}
\def\Fmn{F_{mn}}
\def\bgsC{\mid vac, \downarrow_C\rangle}

\documentclass[twocolumn,showpacs]{revtex4}
\begin{document}

\title{Nonlocal properties of dynamical three-body Casimir-Polder forces}

\author{L. Rizzuto, R. Passante and F. Persico}
\affiliation{Dipartimento di Scienze Fisiche ed Astronomiche
dell'Universit\`{a} degli Studi di Palermo and CNISM, Via
Archirafi 36, I-90123 Palermo, Italy }

\email{roberto.passante@fisica.unipa.it}


\begin{abstract}
We consider the three-body Casimir-Polder interaction between
three atoms during their dynamical self-dressing. We show that the
time-dependent three-body Casimir-Polder interaction energy
displays nonlocal features related to quantum properties of the
electromagnetic field and to the nonlocality of spatial field
correlations. We discuss the measurability of this intriguing
phenomenon and its relation with the usual concept of stationary
three-body forces.
\end{abstract}

\maketitle

One striking  aspect of quantum mechanics is the existence of
nonlocal correlations between spatially separated objects
\cite{H94,POP00}. This might in principle give rise to nonlocal
observable effects, although the possibility of using such
correlations for transmitting superluminal signals has been
investigated in many different frameworks with negative
conclusions \cite{NC00}. Quantum fields display nonlocal
correlations too \cite{CDG89} and it is therefore relevant to
investigate if these correlations have observable consequences. In
this letter we discuss observable manifestations of nonlocality of
quantum electromagnetic fields in dynamical three-body
Casimir-Polder forces between three atoms; in the present context
they arise from nonlocal spatial correlations of the electric
field in the vacuum state during the dynamical self-dressing of
the atoms. The physical basis of our work is that Casimir-Polder
(CP) long-range interatomic interactions are directly related to
the electric field correlations evaluated at the positions of the
atoms, both for two- and many-body components
\cite{PT93,CP97,PPR05,PPR06}. In the case of three atoms, in which
we are mainly interested, the CP potential energy is related to
the electric field correlation at the position of two atoms, {\em
dressed} by the third. In the dynamical (i.e. time-dependent)
case, for example during the self-dressing or the spontaneous
decay of one of the three atoms, spatial field correlations have a
nonlocal evolution and we investigate if this has observable
consequences in the three-body Casimir-Polder interaction between
the three atoms. On the other hand, three-body Casimir-Polder
forces can be also obtained, from different physical
considerations, as the interaction of one atom with the field
fluctuations generated by the other two atoms \cite{PP99}. In a
dynamical situation, field fluctuations are expected to propagate
causally, and thus the interaction with the third atom is not
expected to show nonlocal features.

In this paper we compare in detail results from the two approaches
outlined above, and show that apparently contradictory conclusions
derive from the fact that the two approaches consider two
different, albeit related, interaction energies. Also, our results
suggest that what is usually considered as the three-body energy,
as obtained from the energy shift of the system due to the
atom-field interaction, needs a careful definition from the point
of view of measurement theory. We wish to point out that Casimir
forces in non-equilibrium situations have recently attracted much
interest in the literature \cite{HRS04,APS05}, motivated also by
recent high precision measurements. Recently it has also been
suggested that Casimir-Polder forces may be used as a probe to
investigate quantum entanglement of vacuum fluctuations
\cite{CCPPP06}.

We consider three atoms initially in their bare ground state and
interacting with the radiation field in the vacuum state. We wish
to evaluate their three-body CP interaction during the dynamical
self-dressing of the atoms. In \cite{PPT98} it is shown that this
kind of calculations can be considerably simplified by using an
effective Hamiltonian quadratic in the electric charge which can
be obtained by a unitary transformation from the multipolar
coupling Hamiltonian. This effective Hamiltonian is quadratic in
the field operators and involves the dynamical polarizability
$\alpha_i(k)$ of the atoms
\begin{equation}
H = H_0 - \frac 1 2 \sum_i\sum_{\bk j}\alpha_i(k) \bE(\bk j,\br_i,
t) \cdot \bE(\br_i, t) \label{eq:1}
\end{equation}
where $H_0$ is the free Hamiltonian, $i=A,B,C$ labels the three
atoms and $\bE (\bk j,\br_i, t)$ is the Fourier component of the
electric field operator $\bE (\br_i,t)$ evaluated at the position
$\br_i$ of atom $\it i$. In this coupling scheme the momentum
conjugate to the vector potential is the transverse displacement
field, which outside the atoms coincides with the electric field
(transverse plus longitudinal) \cite{CPP95}. We evaluate the
three-body CP interaction energy from the interaction of one atom
with the time-dependent field emitted by the two other atoms; this
is a generalization of a method already used in a time-independent
situation for three atoms \cite{PP99} and also in time-dependent
cases for the two-body potential \cite{RPP04}. Only terms
involving coordinates and polarizabilities of all atoms are
relevant for the three-body potential. According to our previous
work for the time-independent case, a symmetrization over the role
of the (identical) atoms is required at the end of the calculation
in order to get the three-body potential \cite{PP99}
\begin{equation}
\Delta E_3 = - \frac 13 \sum_i \frac 12 \sum_{\bk j}\alpha_i(k)
\langle \bE(\bk j,\br_i, t) \cdot \bE(\br_i, t) \rangle
\label{eq:2}
\end{equation}
where the average is taken on the dressed vacuum state (we work in
the Heisenberg representation). As mentioned, in stationary cases
this quantity yields the well-known three-body component of the CP
potential as evaluated by straightforward sixth-order perturbation
theory \cite{AZ60}. We perform our calculation evaluating the
interaction of atom C with the dynamical virtual photon cloud
created by atoms A and B. We carry out the calculation in the
Heisenberg representation and use perturbation theory up to
second-order in the atomic polarizabilities (fourth order in the
charge). We first evaluate the electric field operator at point
$\br_C$ due to the pair of atoms A and B, respectively located at
$\br_A$ and $\br_B$, initially in their bare ground-state. At
second-order, the parts of this field operator which are relevant
for the three-body CP interaction are only those which include
contributions coming from both atoms A and B. After some algebra,
we finally get these contributions as
\begin{widetext}
\begin{eqnarray}
E_\ell^{(0)}(\br_C,t)&=& i\left(\frac{2\pi\hbar c}{V}\right)^{1/2}
\sum_{\bk j} k^{1/2} (\ekj)_\ell \left( \akj (t) e^{i \bk \cdot
\br_C} - \akjd (t) e^{i\bk\cdot\br_C} \right)
\label{eq:3}\\
E_\ell^{(1)}(\br_C,t)&=&-i\left(\frac{2\pi\hbar c}{V}\right)^{1/2}
\sum_{\bk
j}k^{1/2}(\ekj)_\ell\alpha_A(k)\Fln^{\beta}\frac{1}{\beta}
\biggl(\akj(0)e^{i\bk\cdot\br_B}e^{-ik(ct-\beta)}
-\akjd(0)e^{-i\bk\cdot\br_B}e^{ik(ct-\beta)}\biggr)\theta(ct-\beta)
\nonumber\\
&\ & + (\alpha\rightleftharpoons\beta; A\rightleftharpoons B)
\label{eq:4}\\
E_\ell^{(2)}(\br_C,t)&=&-i(\frac{2\pi\hbar c}{V})^{1/2} \sum_{\bk
j}k^{1/2}(\ekj)_\ell\alpha_A(k)\alpha_B(k)\Flm^\gamma\Fln^{\beta}\frac{1}{\beta\gamma}
\biggl(\akj(0)e^{i\bk\cdot\br_B}e^{-ik(ct-\beta-\gamma)}
-\akjd(0)e^{-i\bk\cdot\br_B}e^{ik(ct-\beta-\gamma)}\biggr)\nonumber\\
&\ & \times \theta(ct-\beta)\theta(ct-\beta-\gamma)+
(\alpha\rightleftharpoons\beta; A\rightleftharpoons B)
\label{eq:5}
\end{eqnarray}
\end{widetext}
where $\alpha = \mid \br_{BC} \mid = \mid \br_B-\br_C\mid$, $\beta
= \mid \br_{AC} \mid =\mid \br_A-\br_C\mid$ and $\gamma = \mid
\br_{AB} \mid = \mid \br_A-\br_B\mid$; we have also defined the
differential operator $F_{ij}^R =\left( -\nabla^2 \delta_{ij} +
\nabla_i \nabla_j \right)^R$ acting on the variable $\bR$. The
operator $E_\ell^{(0)}(\br_C,t)$ is the free-field operator at
time t, while $E_\ell^{(1)}(\br_C,t)$ and $E_\ell^{(2)}(\br_C,t)$
are the source-dependent contributions to the electric field
operator. The presence of the $\theta$ functions expresses the
causal behaviour of the source electromagnetic field operator. The
field operator in (\ref{eq:4}) is the sum of the contributions
from atoms A and B; in (\ref{eq:5}) only relevant contributions
involving polarizabilities of both atoms have been explicitly
written.

Expressions (\ref{eq:3}-\ref{eq:5}) are now used for evaluating
the quantum average $\langle E_\ell(\bk j,\br_C,t)
E_m(\bk'j',\br_C,t)\rangle$ on the ground state of atom C
appearing in (\ref{eq:2}). Keeping only terms proportional to
$\alpha_A \alpha_B \alpha_C$, we obtain
\begin{widetext}
\begin{eqnarray}
\Delta E_C &=& - \frac 1 2 \sum_{\bk j}\alpha_C(k) \langle \bE(\bk
j,\br_C, t) \cdot \bE(\br_C, t) \rangle = -\frac{2\pi\hbar c}V
\sum_{\bk j} \alpha_A(k)\alpha_B(k)\alpha_C(k)
\nonumber \\
&\times& \Re\Biggl
\{\Flm^\beta\Fln^\alpha\frac{1}{\alpha\beta}(\ekj)_m (\ekj)_n k
e^{i\bk\cdot\br_{AB}}
e^{ik(\beta-\alpha)}\theta(ct-\beta)\theta(ct-\alpha) -
\biggl[\Flm^\beta\Fmn^\gamma\frac{1}{\gamma\beta} (\ekj)_l
(\ekj)_n k e^{-i\bk\cdot\br_{CB}}e^{ik\beta}
\nonumber \\
&\times& \left(\cos k\gamma
-sgn(\gamma-\beta+ct)\frac{e^{-ik\gamma}}{2}-
sgn(\gamma+\beta-ct)\frac{e^{ik\gamma}}{2}\right)\theta(ct-\beta)
+\left(\alpha\rightleftharpoons\beta, A\rightleftharpoons B\right)
\biggr] \biggr\} \label{eq:6}
\end{eqnarray}
\end{widetext}
The physical meaning of this expression is a three-body
contribution to the dynamical CP potential, obtained evaluating
the response of atom C to the fields created by atoms A and B. The
time dependence of the potential is contained in the $sgn$ and
$\theta$ functions. The time-dependent energy (\ref{eq:6}) yields
a time-dependent force on C which in principle can be measured by
a local measurement on atom C. It is non-vanishing only when atom
$C$ is inside the light-cone of both atoms $A$ and $B$, in
agreement with relativistic causality, whatever is the distance
between $A$ and $B$. Indeed, $\Delta E_C$ is different from zero
even if atoms $A$ and $B$ are space-like separated (but both
inside the light cone of atom $C$).

The usual three-body CP potential is however given by the
symmetrized quantity (\ref{eq:2}) \cite{PP99}. Assuming identical
atoms we can permutate their role in (\ref{eq:6}), in order to
obtain the two other contributions in (\ref{eq:2}), that is the
interaction of A with the field generated by B and C ($\Delta
E_A$) and the interaction of B with the field generated by A and C
($\Delta E_B$). We do not report here the final expression of
$\Delta E_3$ because only a few limiting cases are relevant for
our purpose of discussing nonlocal aspects of the CP interaction
energy. In the limit of large times, $\Delta E_3$ reduces to the
well-known stationary expression, as expected, confirming the
validity of our approach. If each atom is causally disconnected
from the other two, $\Delta E_3$ vanishes also, in agreement with
relativistic causality. The most interesting feature is however
for configurations such that two atoms are inside the light-cone
of the third one, but they have a space-like separation. An
example is when $\alpha ,\beta < ct$ and $\gamma > ct$, with A and
B inside the light cone of C but causally disconnected from each
other. In this case we find $\Delta E_A = \Delta E_B =0$ and the
symmetrized potential reduces to
\begin{eqnarray}
\Delta E_3 &=& - \frac{2\pi \hbar c}{6V}\Flm^\beta\Fln^\alpha
\frac{1}{\alpha\beta} \Re \sum_{\bk j} (\ekj )_m (\ekj )_n k
\nonumber\\
&\times& \alpha_A(k)\alpha_B(k)\alpha_C(k) e^{i\bk \cdot \br_{AB}}
e^{ik(\beta-\alpha)}
\label{eq:7}
\end{eqnarray}
which is different from zero. Thus we find that some nonlocal
aspects are present in the time dependence of this potential
because we obtain a nonvanishing three-body potential even if two
of the three atoms are not causally connected. This is not trivial
because a three-body interaction involves the presence of all
three atoms, and it is expected to be nonvanishing only if each
atom is aware of the presence of the other two. This point is
clearly related to an inherent nonlocal nature of three-body
potentials and does not raise any causality violation issue. In
fact, the potential (\ref{eq:7}) is related to the {\em local}
interaction of atom C with the virtual field dressing atoms A and
B. As shown in Eq.(\ref{eq:6}), this contribution behaves
causally, since a change in the field dressing A and B is felt by
C only after the causality time $t>\alpha /c, \beta /c$. We stress
that in this approach the three-body potential has been obtained
by evaluation of the local interaction of each atom with the
time-dependent field fluctuations dressing the other two atoms.

We now address the same problem of nonlocality of three-body CP
interactions from a different point of view, already mentioned in
the introduction. A relevant question is if the nonlocal
properties of the electromagnetic field induce observable nonlocal
features in the time-dependent CP interaction. This is expected
because the CP interaction between two or more atoms depends on
vacuum field correlations evaluated at the atomic positions, and
it is known that field correlations behave nonlocally
\cite{PT93,PPR06,PPR03,CPPP03}. Let us consider an atom (say C)
initially in its bare ground-state and interacting with the
electromagnetic field in its vacuum state. This system can be
described by the effective Hamiltonian (\ref{eq:2}) with $i=C$. We
first evaluate the average value of the equal-time spatial
correlation of the electric field on the initial state $\bgsC$ at
two different points $\br_A$ and $\br_B$, $\langle E_\ell (\br_A
,t) E_m (\br_B,t) \rangle$, during the dynamical self-dressing of
atom C. In the Heisenberg representation and up to first order in
$\alpha_C$, the correlation function is $\langle E_\ell
(\br_A,t)E_m (\br_B,t)\rangle = \langle
E_\ell^{(0)}(\br_A,t)E_m^{(0)}(\br_B,t)\rangle + \langle
E_\ell^{(0)}(\br_A,t)E_m^{(1)}(\br_B,t)\rangle + \langle
E_\ell^{(1)}(\br_A,t)E_m^{(0)}(\br_B,t)\rangle$ (the last two
terms contain source contributions, related to the three-body
potential) . Nontrivial results are obtained in this case which
can be summarized as follows (more details can be found in
\cite{PPR06}). (i) If both points $\br_A$ and $\br_B$ are outside
the causality sphere of atom C, the correlation function reduces
to the free-field correlation; (ii) if $\br_A$ and $\br_B$ are
both inside the causality sphere of C, the correlation function is
modified by the presence of atom C; (iii) if just one of the two
points $\br_A$ and $\br_B$ is inside the causality sphere of C,
the spatial field correlation is still modified by C. The latter
point is remarkable and emphasizes the nonlocal features of field
correlations. As an example, let us consider the case $\alpha <
ct$ and $\beta > ct$: in this case the correlation function is
modified by C. Nevertheless the atom in $\br_A$ does not yet
``know" of the presence of the atom in $\br_C$. Thus source
contributions to field correlations exhibit nonlocal properties,
similarly to free-field correlations. It seems now important to
inquire whether this nonlocal behaviour induces observable effects
in the time-dependent three-body CP interaction between two
ground-state atoms, A and B, located at $\br_A$ and $\br_B$
respectively. As we have shown in \cite{PPR06}, the three-body
interaction energy between A and B in the presence of C can be
obtained in the following form
\begin{eqnarray}
&\ &\Delta E_C(A,B) = \sum_{\bk j, \bk'
j'}\alpha_{A}(k)\alpha_{B}(k')
\nonumber\\
&\ &\times \langle E_\ell(\bk j,\br_A,t)E_m(\bk'
j',\br_B,t)\rangle V_{lm}(k,\br_{A}, k',\br_{B}) \label{eq:8}
\end{eqnarray}
where $V_{lm}(k,\br_{A}, k',\br_{B})$ is the usual classical
potential tensor between oscillating dipoles \cite{CP97} and only
relevant source contributions are included in the expectation
value of the field correlation on the state $\bgsC$.

Let us now consider the specific configuration $\alpha , \beta >
ct$ and $\gamma < ct$. This means that both A and B are outside
the light cone of C, but they are inside the light cone of each
other. The interaction energy $\Delta E_C(A,B)$ between A and B is
then
\begin{eqnarray}
&\ &\Delta E_C(A,B) =- \frac{\hbar c}{16\pi}\Flm^\beta\Fln^\alpha
\Fmn^\gamma\frac{1}{\alpha\beta\gamma}\int_0^{\infty}du \alpha_A(iu)\nonumber\\
&\ & \times\alpha_B(iu)\alpha_C(iu)
(1-sgn(\alpha-\gamma-ct))(e^{-u(\beta-\alpha+\gamma)}
\nonumber\\
&\ & +e^{-u(\alpha+\beta-\gamma)})+(\alpha\rightleftharpoons\beta)
\label{eq:9}
\end{eqnarray}
The essential point resulting from eq. (\ref{eq:9}) is that, in
contrast with the conclusions drawn from (\ref{eq:6}), time
intervals exist for which $\Delta E_C(A,B)$ is nonvanishing: this
happens when $ct<\alpha <\gamma +ct$ and/or $ct<\beta <\gamma
+ct$. We stress that atoms A and B are outside the light come of
C: nonetheless their CP interaction energy is affected by atom C,
displaying nonlocality in the interaction energy. Thus the
nonlocal features of field correlations discussed above induce an
observable nonlocal behaviour of the time-dependent CP potential.
We can also consider the symmetrized form of the three-body CP
potential, obtained by averaging over the interaction energy of
any pairs of atoms in the presence of the third one. This is the
quantity that in the limit of large times yields the same
stationary potential obtained in a time-independent approach
\cite{CP97}. It is easy to show from (\ref{eq:9}) that also the
symmetrized interaction energy has nonlocal features. Thus we
conclude that the nonlocal properties of the quantum
electromagnetic field may induce observable effects in dynamical
three-body interatomic interactions.

The contrast between (\ref{eq:6}) and (\ref{eq:9}) can be resolved
by remarking that the physical meaning of the interaction energy
$\Delta E_C(A,B)$ in (\ref{eq:9}) is very different from the
meaning of the interaction energy $\Delta E_C$ given by eq.
(\ref{eq:6}). In fact the former is the interaction energy between
two atoms (A and B) in the presence of the third one (C); its
measurements therefore necessitates {\em correlated} measurements
on A and B and not a single {\em local} measurement as for the
latter. In conclusion, the various different quantities we have
evaluated, all related to a three-body Casimir-Polder interaction
energy, represent meaningful quantities differing in how they are
measured: a single local measurement on one atom or correlated
measurements on two atoms. This consideration makes our result
showing sharp nonlocal features in (\ref{eq:9}) less surprising.
In addition, the usual three-body potential (as inferred from
stationary perturbative calculations) involves a symmetrization of
the quantities discussed above on the role of the three atoms, and
therefore it is not equally clear how it can be measured. In
contrast, the physical meaning of the interaction energies in
(\ref{eq:6}) and (\ref{eq:9}) is clear as well as how they can be
measured. Finally, we wish to stress that the current interest in
Casimir-Polder forces is not only academic, because Casimir-Polder
forces involving atoms has been recently measured with precision
\cite{SBCSH93,PC05,HOMC05}. It is therefore worth a deep
investigation of their properties both in stationary and dynamical
situations.

\begin{acknowledgments}
This work has been partially supported by the bilateral
Italian-Belgian project on ``Casimir-Polder forces, Casimir effect
and their fluctuations" and by the bilateral Italian-Japanese
project 15C1 on ``Quantum Information and Computation" of the
Italian Ministry for Foreign Affairs. Partial support by Ministero
dell'Universit\`{a} e della Ricerca Scientifica e Tecnologica and
by Comitato Regionale di Ricerche Nucleari e di Struttura della
Materia is also acknowledged.
\end{acknowledgments}

\end{document}